\documentclass[twocolumn,prl,amsmath,amssymb]{revtex4}

\usepackage{graphicx}
\usepackage[utf8]{inputenc}

\begin{document}

\title{Comment on ``Correlation between Dynamic Heterogeneity and Medium-Range Order in Two-Dimensional Glass-Forming Liquids''}

\author{François Sausset}
\email{sausset@lptmc.jussieu.fr}
\affiliation{Laboratoire de Physique Théorique de la Matière Condensée, Université Pierre et Marie Curie - Paris 6, UMR CNRS 7600, 4 place Jussieu, 75252 Paris Cedex 05, France} 

\author{Gilles Tarjus}
\email{tarjus@lptl.jussieu.fr}
\affiliation{Laboratoire de Physique Théorique de la Matière Condensée, Université Pierre et Marie Curie - Paris 6, UMR CNRS 7600, 4 place Jussieu, 75252 Paris Cedex 05, France} 

\maketitle

In a recent letter, Kawasaki \textit{et al.} \cite{Kawasaki:2007} study glass formation in a $2$-dimensional ($2D$) model of polydisperse repulsive disks. They give numerical evidence for a direct relation between slowing down of the relaxation, extension of some medium-range order, and development of dynamic heterogeneities. We disagree with the authors' interpretation on two points which we believe to be potentially important for understanding the glass transition.

\begin{figure}
	\includegraphics[scale=0.3]{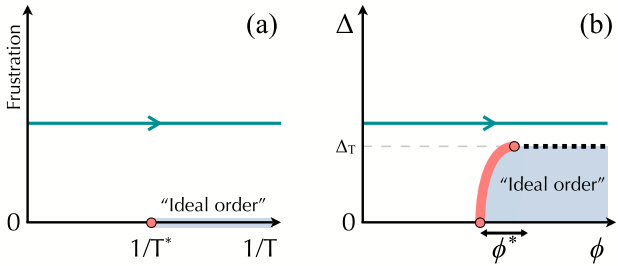}
	\caption{\label{fig:phasediagram}Schematic diagrams: (a) frustration/temperature for a $3D$ system; (b) polydispersity/density for a $2D$ polydisperse system. In both cases the (critical or weakly first order) transition to a phase with long-range extension of the locally preferred structure of the liquid (``ideal order'') is avoided in the range of interest for glass formation: in nonzero frustration in (a) and for $\Delta$ larger than $\Delta_{T}$ in (b) (full straight line with arrow). In (a)  the transition occurs at a single $T^*$ whereas in (b) a continuous transition occurs for a range of density (shown by the arrow around $\phi^*$) with the transition becoming strongly first order (and irrelevant) in the almost horizontal part shown by a short-dashed line. (The phase diagram of (b) is adapted from \cite{Sadr-Lahijany:1997,Pronk:2004}.)}
\end{figure}

First, contrary to what suggested in \cite{Kawasaki:2007}, the order that develops with increasing density when polydispersity is large enough for thwarting crystallization ($\Delta \gtrsim 9\%$) should not be contrasted with icosahedral order in $3D$ (metallic) liquids and glasses and, for that reason, is not appropriately  characterized as ``crystalline''. The local order detected by Kawasaki \textit{et al.} is hexatic/hexagonal, which represents the locally preferred order of the \textit{liquid}, just like icosahedral order in $3D$. When polydispersity is larger than a threshold value $\Delta_T$, this liquid local order cannot extend over the whole sample. This frustration is akin to the geometric frustration that prevents tiling of $3D$ Euclidean space by icosahedral order \cite{Tarjus:2005,Kivelson:1995}. The parallel between the two situations is sketched in Fig. \ref{fig:phasediagram}. In the presence of frustration, crystalline order is either absent or different from the liquid locally preferred order.

Secondly, the observations reported in \cite{Kawasaki:2007} nicely fit in the frustration-limited domain (FLD) approach \cite{Tarjus:2005,Kivelson:1995} based on the above frustration description: for large enough polydispersity (\textit{i.e.} in the presence of frustration), the continuous or weakly first-order transition occuring in the narrow range between $\phi_I(\Delta=0)$ and $\phi_I(\Delta_T)$ is avoided (see Fig. \ref{fig:phasediagram}) and marks the onset of cooperative slow dynamics and dynamic heterogeneity, both being associated with the ``frustration-limited domains'' (clearly visible in Fig.2a of \cite{Kawasaki:2007}). As stressed in \cite{Tarjus:2005,Kivelson:1995}, the more frustrated (here, the larger the polydispersity), the less collective the behavior, hence the less ``fragile'' the liquid, as indeed observed in \cite{Kawasaki:2007}. Within the FLD approach there is no need to invoke an unreachable ideal glass transition point $\phi_0$ to explain the phenomenology. (This turns out to be relevant in the case of $2D$ polydisperse systems of repulsive disks, for which evidence has been given that there is no such ideal transition \cite{Santen:2000}.) Our interpretation is rather that the avoided singularity leads, beyond some crossover $\phi^*$, to a scaling behavior according to which the relaxation time $\tau_\alpha$ and the size $L^*$ of the FLD's with extended locally preferred liquid order (here hexatic/hexagonal order) behave as
\begin{equation*}
log \left ( \frac{\tau_\alpha }{\tau_0} \right ) \sim L^{*\, \theta}, \; L^*\sim B \left (\frac{\phi - \phi^*}{\phi^*} \right )^x + \mathrm{constant},
\end{equation*}
with $B$ decreasing with increasing frustration and $\theta$, $x$ two exponents whose value depends on the space dimensionality. In the notations of \cite{Kawasaki:2007}, $L^*$ identifies with $\xi$ (hence with $\xi_6$) and, apparently, $\theta \simeq1$. We checked that the data of Fig.4a of \cite{Kawasaki:2007} can indeed be fitted by the above formula with $x\simeq 3$, $\phi^* \simeq \phi_I$, and $B$ monotonically decreasing as polydispersity increases. This description therefore emphasizes the role of the avoided transition (in zero or very small polydispersity) in triggering the slowing down of relaxation.

\bibliography{Bib}

\end{document}